\documentclass{caosp306}

\usepackage{graphicx}

\usepackage{natbib}
\articleNo{A08}
\pubyear{2019}
\volume{49}
\volnumber{2}
\firstpage{1}
\received{November 5, 2018}
\accepted{February 1, 2019}

\begin{document}

\hauthor{M.\,Skarka and P.\,Kab\'{a}th}

\title{Synergy between professional and amateur astronomers}

\author{
	M.\,Skarka\inst{1,2} 
	\and
	P.\,Kab\'{a}th\inst{2}
       }

\institute{
	   Department of Theoretical Physics and Astrophysics, Masaryk University, Kotl\'{a}\v{r}sk\'{a} 2, CZ-611\,37 Brno, Czech Republic \email{maska@physics.muni.cz}
         \and
           Astronomical Institue, Czech Academy of Sciences, Fri\v{c}ova 298, 251\,65 Ond\v{r}ejov, Czech Republic
           }

\date{November 5, 2018}

\maketitle

\begin{abstract}
Since the CCD technique became financially reachable for amateur astronomers, they can cover topics of professional science. Mainly in the time-domain astronomy, such as variable star research, their help is invaluable. We focus on a cooperation between amateur and professional astronomers in the Czech Republic, give some examples of successful projects and propose new programs that can benefit from such cooperation and bring high-quality results.
\end{abstract}

\section{Introduction}\label{Sect:Introduction}

Even in the time of large sky surveys the new multicolour photometric and mainly spectroscopic data are desirable. Without the ground-based follow-up observations, the space missions such as {\it Kepler} \citep{borucki2010} cannot provide us with the full spectrum of information. Because getting the data for many objects is very time demanding and observing time at professional observatories is often very restricted, the incorporation of amateur astronomers (AMs) to the professional projects could be very helpful. 

Mainly in time-domain astronomy (for example, multicolour photometry of variable stars), AMs under the coordination of professional astronomers (PROs) can contribute significantly. By using telescopes with 20-30cm in diameter, which are common among AMs, with proper equipment the AMs can get single-point precision down to several mmag. Because of the technical and data-processing demands, spectroscopy is not common among AMs, although some exceptions exist.

\section{Variable star research in the Czech Republic}\label{Sect:VSresearch}

Observation of variable stars has a long tradition in the Czech Republic and the relation between AMs and PROs is very tight. The AMs are united in the Variable Stars and Exoplanet Section (VSES) of the Czech Astronomical Society, which was founded in 1924 \citep{skarka2015}. For a long time, the main topic was the observation of minima of eclipsing binary stars. However, the AMs currently observe also exoplanetary transits, eruptive and pulsating stars.

The VSES maintains two important databases: the Exoplanet Transit Database \citep[ETD,][]{poddany2010} and the O-C gate \citep{paschke2006a}, which are used by astronomers around the world. The VSES also provides a possibility to publish the results of AMs observations -- the Open European Journal on Variable stars \citep[OEJV,][]{paschke2006b}.

The Czech AMs participated in very nice projects and discovered interesting binaries. For example, \citet{cagas2012} studied the double eclipsing binary with periods near 3:2 ratio, AMs contributed to the study of multiple stellar systems \citep{zasche2017} and eruptive stars \citep{smelcer2017}, an amateur observer allowed for the precise description of modulation properties of Z CVn \citep{skarka2018}.

\section{Possible projects with AMs in the era of TESS}\label{Sect:Projects}

The Transiting Exoplanet Survey Satellite (TESS) was launched on April 18, 2018. This mission is designed to scan the whole sky and search for exoplanetary transits \citep{ricker2015}. However, the satellite will be extremely useful in discovering and monitoring variable stars. Due to the observing strategy, most of the sky will be observed only for 27 days with 30-min cadence\footnote{About 200\,000 stars will be monitored also in a 2-min mode.} in one broadband filter (spatial resolution 21\,arcsec/px). This opens a great opportunity for the multicolour photometric and spectroscopic follow-up observations. Again, mainly in photometric observations, the AMs can contribute significantly. 

\subsection{RR Lyrae and Cepheid observations}\label{Subsect:RRLCep}

In the investigation of long-term phenomena in RR Lyrae and Cepheid type stars, it is extremely important to have a regular sampling and the longest possible time base of the data. We demonstrate this need on the artificially generated data of a sample RR Lyrae light curve with modulation known as the Blazhko effect \citep{blazhko1907}.

\begin{figure}
\centerline{\includegraphics[width=0.95\textwidth,clip=]{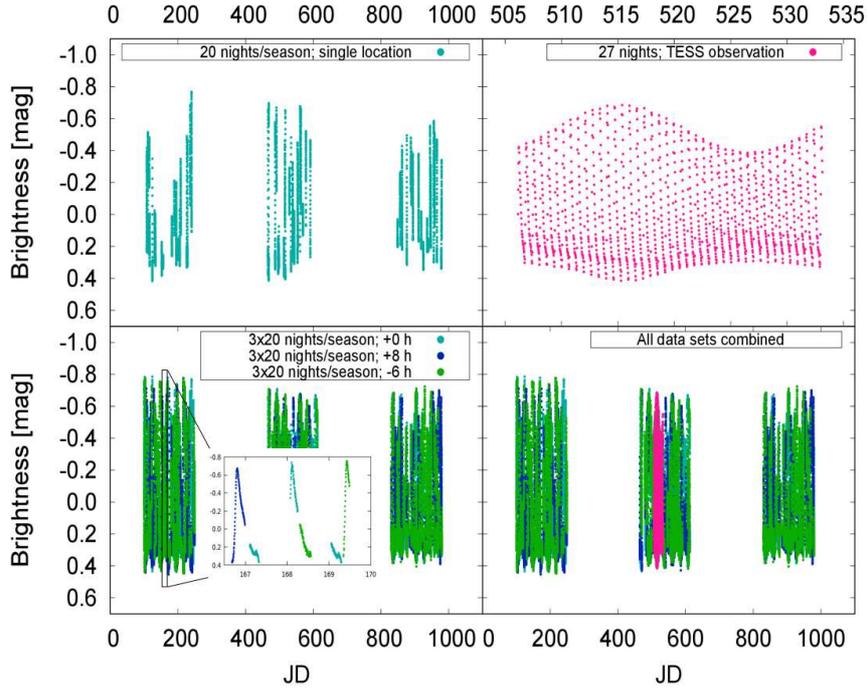}}
\caption{The simulated data. See the text for the description.}
\label{Fig:Data}
\end{figure}

We simulated the 27-days TESS observations with 30-minutes cadence (see the top right-hand panel of Fig.\,\ref{Fig:Data}) by using the mathematical description of a modulated RR Lyrae star by adding the white noise of 0.001 mag, which roughly corresponds to the expected accuracy for a 12.5-mag star \citep{ricker2015}. 

The ground-based observations were simulated similarly, but by adding 0.01-0.05-mag white noise to the observations (top left panel of Fig.\,\ref{Fig:Data}), which is typical of 20-30\,cm telescopes with CCD in various filters. We considered three locations (0, +8 hours, $-$6 hours) and cadence 4 minutes to simulate three ground-based observatories. The data cover 3 seasons in a random range up to 120 days, with the random start of the observations (within the local nights, of course) and the random length of the night to simulate the weather conditions (the bottom panels of Fig. \ref{Fig:Data}).

\begin{figure}
\centerline{\includegraphics[width=0.9\textwidth,clip=]{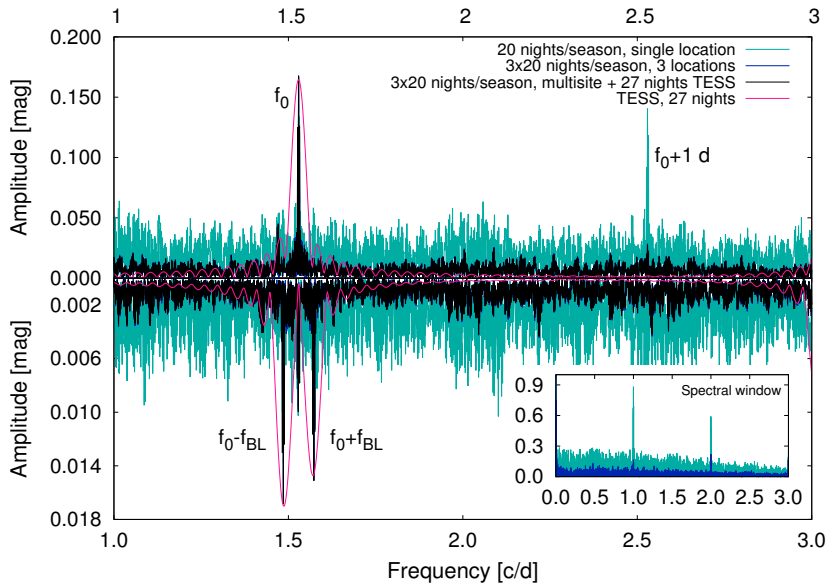}}
\caption{Frequency spectra of the artificially generated data (top panel) and the residuals after removal of the basic pulsation frequency $f_{0}$ (bottom panel). On the bottom panel, the peaks $f_{0}\pm f_{\rm BL}$ are generated by the Blazhko effect. The spectral window is shown in the detail.}
\label{Fig:FrekSpec}
\end{figure}

The resulting frequency spectra of the TESS and combined data are shown in Fig.\,\ref{Fig:FrekSpec}. The spectrum of TESS data with a 27-days long time base has very broad peaks and show no apparent aliases. On the other hand, the spectrum of the ground-based data from one observing base suffers from strong daily aliases. The aliases are significantly reduced in the data from three observing sites and almost disappear in the combined dataset shown by the black line. The longer the database, the better the frequency resolution. For the 3-year combined data set the resolution is $1/(3\times 365)\approx0.001$\,cd$^{-1}$, which is 40 times better than for the TESS observations. This clearly shows the advantages of a multi-site observing campaign and the time demands of such campaign. Therefore, the cooperation with AMs is considered in the research of RR Lyrae and Cepheid stars observed by TESS.

\newpage
\vspace*{-1cm}

\subsection{Follow-up of exoplanets}\label{Subsect:Exo}
Amateur observers can also significantly contribute to the follow-up observations of exoplanetary candidates. Due to short observing runs of TESS, the transits detected from the ground will help to better estimate the orbital parameters. The observations will also help to detect peculiarities in the transits and identify blends that were unresolved by TESS. Detection of secondary transits (eclipse minima) will help to identify binaries that were misclassified as exoplanets. This clearly shows the need for a large amount of observing time that would be impossible to get on professional observatories. In addition, the multi-site observations will increase the amount of gathered data.

To unambiguously constrain the real planetary nature of the candidate, spectroscopic radial velocity observations will be necessary. For that, 2-m Perek telescope in Ondrejov, Czech Republic can be used (see Kab\'{a}th et al. 2019, these proceedings for more details).

\section{Conclusions}\label{Sect:Conclusions}

We discussed the possible involvement of amateurs in professional projects. Their help can be especially important in the TESS follow-up observations. We propose two projects, one focused on RR Lyrae and Cepheid stars, the second on exoplanetary candidates, which will be organized in the near future in collaboration with the home institutes and in combination with the instruments available there (the 2-m telescope in Ond\v{r}ejov (spectroscopy) and a 0.6-m telescope (photometry) in Brno).


\acknowledgements
MS acknowledges financial support of Postdoc@MUNI project CZ.02.2.69/0.0/0.0/16 027/0008360. PK acknowledges GACR international grant 17-01752J.

\bibliography{demo_caosp306}

\begin{thebibliography}{11}
\expandafter\ifx\csname natexlab\endcsname\relax\def\natexlab#1{#1}\fi

\bibitem[{{Bla{\v z}ko}(1907)}]{blazhko1907}
{Bla{\v z}ko}, S., {Mitteilung {\"u}ber ver{\"a}nderliche Sterne}. 1907, {\it
  Astronomische Nachrichten}, {\bf 175}, 325, DOI: 10.1002/asna.19071752002

\bibitem[{{Borucki} {et~al.}(2010){Borucki}, {Koch}, {Basri}, {Batalha},
  {Brown}, {Caldwell}, {Caldwell}, {Christensen-Dalsgaard}, {Cochran},
  {DeVore}, {Dunham}, {Dupree}, {Gautier}, {Geary}, {Gilliland}, {Gould},
  {Howell}, {Jenkins}, {Kondo}, {Latham}, {Marcy}, {Meibom}, {Kjeldsen},
  {Lissauer}, {Monet}, {Morrison}, {Sasselov}, {Tarter}, {Boss}, {Brownlee},
  {Owen}, {Buzasi}, {Charbonneau}, {Doyle}, {Fortney}, {Ford}, {Holman},
  {Seager}, {Steffen}, {Welsh}, {Rowe}, {Anderson}, {Buchhave}, {Ciardi},
  {Walkowicz}, {Sherry}, {Horch}, {Isaacson}, {Everett}, {Fischer}, {Torres},
  {Johnson}, {Endl}, {MacQueen}, {Bryson}, {Dotson}, {Haas}, {Kolodziejczak},
  {Van Cleve}, {Chandrasekaran}, {Twicken}, {Quintana}, {Clarke}, {Allen},
  {Li}, {Wu}, {Tenenbaum}, {Verner}, {Bruhweiler}, {Barnes}, \&
  {Prsa}}]{borucki2010}
{Borucki}, W.~J., {Koch}, D., {Basri}, G., {et~al.}, {Kepler Planet-Detection
  Mission: Introduction and First Results}. 2010, {\it Science}, {\bf 327},
  977, DOI: 10.1126/science.1185402

\bibitem[{{Caga{\v s}} \& {Pejcha}(2012)}]{cagas2012}
{Caga{\v s}}, P. \& {Pejcha}, O., {Discovery of a double eclipsing binary with
  periods near a 3:2 ratio}. 2012, {\it \aap}, {\bf 544}, L3, DOI:
  10.1051/0004-6361/201219815

\bibitem[{{Paschke} \& {Brat}(2006{\natexlab{a}})}]{paschke2006a}
{Paschke}, A. \& {Brat}, L., {O-C Gateway, a Collection of Minima Timings}.
  2006{\natexlab{a}}, {\it Open European Journal on Variable Stars}, {\bf 23},
  13

\bibitem[{{Paschke} \& {Brat}(2006{\natexlab{b}})}]{paschke2006b}
{Paschke}, A. \& {Brat}, L., {Open European Journal on Variable Stars}.
  2006{\natexlab{b}}, {\it Open European Journal on Variable Stars}, {\bf 23},
  15

\bibitem[{{Poddan{\'y}} {et~al.}(2010){Poddan{\'y}}, {Br{\'a}t}, \&
  {Pejcha}}]{poddany2010}
{Poddan{\'y}}, S., {Br{\'a}t}, L., \& {Pejcha}, O., {Exoplanet Transit
  Database. Reduction and processing of the photometric data of exoplanet
  transits}. 2010, {\it New Astronomy}, {\bf 15}, 297, DOI:
  10.1016/j.newast.2009.09.001

\bibitem[{{Ricker} {et~al.}(2015){Ricker}, {Winn}, {Vanderspek}, {Latham},
  {Bakos}, {Bean}, {Berta-Thompson}, {Brown}, {Buchhave}, {Butler}, {Butler},
  {Chaplin}, {Charbonneau}, {Christensen-Dalsgaard}, {Clampin}, {Deming},
  {Doty}, {De Lee}, {Dressing}, {Dunham}, {Endl}, {Fressin}, {Ge}, {Henning},
  {Holman}, {Howard}, {Ida}, {Jenkins}, {Jernigan}, {Johnson}, {Kaltenegger},
  {Kawai}, {Kjeldsen}, {Laughlin}, {Levine}, {Lin}, {Lissauer}, {MacQueen},
  {Marcy}, {McCullough}, {Morton}, {Narita}, {Paegert}, {Palle}, {Pepe},
  {Pepper}, {Quirrenbach}, {Rinehart}, {Sasselov}, {Sato}, {Seager},
  {Sozzetti}, {Stassun}, {Sullivan}, {Szentgyorgyi}, {Torres}, {Udry}, \&
  {Villasenor}}]{ricker2015}
{Ricker}, G.~R., {Winn}, J.~N., {Vanderspek}, R., {et~al.}, {Transiting
  Exoplanet Survey Satellite (TESS)}. 2015, {\it Journal of Astronomical
  Telescopes, Instruments, and Systems}, {\bf 1}, 014003, DOI:
  10.1117/1.JATIS.1.1.014003

\bibitem[{{Skarka} {et~al.}(2018){Skarka}, {Li{\v s}ka}, {D{\v r}ev{\v
  e}n{\'y}}, {Guggenberger}, {S{\'o}dor}, {Barnes}, \&
  {Kolenberg}}]{skarka2018}
{Skarka}, M., {Li{\v s}ka}, J., {D{\v r}ev{\v e}n{\'y}}, R., {et~al.}, {A
  cautionary tale of interpreting O-C diagrams: period instability in a
  classical RR Lyr star Z CVn mimicking as a distant companion}. 2018, {\it
  \mnras}, {\bf 474}, 824, DOI: 10.1093/mnras/stx2737

\bibitem[{{Skarka} {et~al.}(2015){Skarka}, {Li{\v s}ka}, {{\v S}melcer}, \&
  {Br{\'a}t}}]{skarka2015}
{Skarka}, M., {Li{\v s}ka}, J., {{\v S}melcer}, L., \& {Br{\'a}t}, L.,
  {Variable Star and Exoplanet Section of the Czech Astronomical Society}.
  2015, in Astronomical Society of the Pacific Conference Series, Vol. {\bf
  496}, {\it Living Together: Planets, Host Stars and Binaries}, ed. S.~M.
  {Rucinski}, G.~{Torres}, \& M.~{Zejda}, 307

\bibitem[{{{\v S}melcer} {et~al.}(2017){{\v S}melcer}, {Wolf}, {Ku{\v
  c}{\'a}kov{\'a}}, {B{\'{\i}}lek}, {Dubovsk{\'y}}, {Ho{\v n}kov{\'a}}, \&
  {Vra{\v s}til}}]{smelcer2017}
{{\v S}melcer}, L., {Wolf}, M., {Ku{\v c}{\'a}kov{\'a}}, H., {et~al.}, {Flare
  activity on low-mass eclipsing binary GJ 3236}. 2017, {\it \mnras}, {\bf
  466}, 2542, DOI: 10.1093/mnras/stw3179

\bibitem[{{Zasche} {et~al.}(2017){Zasche}, {Jury{\v s}ek}, {Nemravov{\'a}},
  {Uhla{\v r}}, {Svoboda}, {Wolf}, {Ho{\v n}kov{\'a}}, {Ma{\v s}ek}, {Prouza},
  {{\v C}echura}, {Kor{\v c}{\'a}kov{\'a}}, \& {{\v S}lechta}}]{zasche2017}
{Zasche}, P., {Jury{\v s}ek}, J., {Nemravov{\'a}}, J., {et~al.}, {V773 Cas, QS
  Aql, and BR Ind: Eclipsing Binaries as Parts of Multiple Systems}. 2017, {\it
  \aj}, {\bf 153}, 36, DOI: 10.3847/1538-3881/153/1/36

\end{thebibliography}

\end{document}